\documentclass[12pt,manuscript]{emulateapj}

\newcommand{\perbeam}{\,beam$^{-1}$}
\newcommand{\source}{4U~1957+11}
\def\simlt{\mathrel{\rlap{\lower 3pt\hbox{$\sim$}}
        \raise 2.0pt\hbox{$<$}}}
\def\simgt{\mathrel{\rlap{\lower 3pt\hbox{$\sim$}}
        \raise 2.0pt\hbox{$>$}}}

\shorttitle{An ultradeep observation of a soft state BH}
\shortauthors{Russell et al.}

\begin{document}


\title{Testing the jet quenching paradigm with an ultradeep observation of a steadily soft state black hole}


\author{D. M. Russell\altaffilmark{1}, J. C. A. Miller-Jones\altaffilmark{2}, T. J. Maccarone\altaffilmark{3}, Y. J. Yang\altaffilmark{1}, R. P. Fender\altaffilmark{3} and F. Lewis\altaffilmark{4,5}}
\altaffiltext{1}{Astronomical Institute `Anton Pannekoek', University of Amsterdam, P.O. Box 94249, 1090 GE Amsterdam, the Netherlands; d.m.russell,y.j.yang@uva.nl}
\altaffiltext{2}{International Center for Radio Astronomy Research -- Curtin University, GPO Box U1987, Perth, WA 6845, Australia; james.miller-jones@curtin.edu.au}
\altaffiltext{3}{School of Physics and Astronomy, University of Southampton, Southampton SO17 1BJ, UK; t.j.maccarone,r.fender@soton.ac.uk}
\altaffiltext{4}{Faulkes Telescope Project, University of Glamorgan, Pontypridd CF37 1DL, UK; fraser.lewis@faulkes-telescope.com}
\altaffiltext{5}{Department of Physics and Astronomy, The Open University, Walton Hall, Milton Keynes, MK7 6AA, UK}




\begin{abstract}
We present ultradeep radio observations with the Expanded Very Large Array of \source, a Galactic black hole candidate X-ray binary known to exist in a persistent soft X-ray state. We derive a stringent upper limit of 11.4 $\mu$Jy beam$^{-1}$ (3$\sigma$) at 5--7 GHz, which provides the most rigorous upper limit to date on the presence of jets in a soft state black hole X-ray binary. X-ray, UV and optical fluxes obtained within a few weeks of the radio data can be explained by thermal emission from the disk. At this X-ray luminosity, a hard state black hole X-ray binary that follows the established empirical radio--X-ray correlation would be at least 330--810 times brighter at radio frequencies, depending on the distance to \source. This jet quenching of $> 2.5$ orders of magnitude is greater than some models predict, and implies the jets are prevented from being launched altogether in the soft state. \source\ is also more than one order of magnitude fainter than the faintest of the `radio-quiet' population of hard state black holes. In addition, we show that on average, soft state stellar-mass BHs probably have fainter jets than most active galactic nuclei in a state equivalent to the soft state. These results have implications for the conditions required for powerful, relativistic jets to form, and provide a new empirical constraint for time- and accretion mode-dependent jet models, furthering our understanding of jet production and accretion onto BHs.
\end{abstract}





\section{Introduction}

It is now realized that a common consequence of the accretion of matter onto a compact object is the formation of fast, powerful jets. For accreting black holes (BHs) the jets can be relativistic, focusing a large fraction of the accretion energy into the collimated outflows \citep*[e.g.][]{ree84,millfn06,russet07} but only in some accretion states. Specifically, a `unified model' has been proposed for black hole candidate X-ray binaries (BHXBs; stellar-mass BHs accreting from a companion star) linking the jet properties to the behavior of the inflow. The model is based on X-ray luminosity/spectral hysteresis that is seen during BHXB outbursts \citep{miyaet95,maccco03} and can explain the interplay between the radio jet (outflow) properties and X-ray (radiation usually from the inflow) properties of BHXBs based on compilations of data from many sources \citep*[e.g.][]{fendet09}. 
Different X-ray states in BHXBs, and correlated radio jet behavior are likely to be analogous to different \emph{classes} of active galactic nuclei \citep*[AGN; e.g.][]{pounet95,maccet03,fend09}. Recent work \citep*[e.g.][]{corbet00,gallet03,falcet04,gultet09} has provided good observational evidence that empirical correlations can be found to explain jet production in black holes across a large range of masses, indicating 
a common jet production process. As such, results from disk--jet coupling studies of BHXBs can be inferred to apply also to AGN.


\begin{table*}
\begin{center}
\caption{Log of observations.}
\begin{tabular}{lllll}
\tableline\tableline
Date & MJD & Telescope & Instrument & Exposure times/sec (bandpass) \\
\tableline
2010-11-04&55504.98&EVLA &C-band& 1614 (4.868--7.312 GHz)\\
2010-11-16&55516.22&Swift&XRT &1699 (0.3--10 keV) \\
          &        &     &UVOT&140 ($v$, $b$, $u$), 279 ($uvw1$), 385 ($uvm2$), 559 ($uvw2$) \\
2010-11-19&55519.03&Swift&XRT &1610 (0.3--10 keV) \\
          &        &     &UVOT&135 ($v$, $b$, $u$), 269 ($uvw1$), 344 ($uvm2$), 539 ($uvw2$) \\
2010-11-22&55522.91&Swift&XRT &1649 (0.3--10 keV) \\
          &        &     &UVOT&135 ($v$, $b$, $u$), 269 ($uvw1$), 380 ($uvm2$), 539 ($uvw2$) \\
2010-11-23&55523.22&FTN  &EM01&219 (B), 160 (V), 221 (R), 200 (i') \\
2010-11-25&55525.58&Swift&XRT &1624 (0.3--10 keV) \\
          &        &     &UVOT&135 ($v$, $b$, $u$), 269 ($uvw1$), 354 ($uvm2$), 539 ($uvw2$) \\
2010-12-09&55539.94&EVLA &C-band& 7272 (4.868--7.312 GHz)\\
\tableline
\end{tabular}
\newline
\end{center}
\end{table*}

One of the major (and first) couplings observed was a suppression of the radio emission from the jet when BHXBs are in the soft state \citep[e.g.][]{tanaet72} compared to the hard state. This has now been confirmed; the jet is dramatically suppressed in the soft state \citep[e.g.][]{gallet03,fendet09}. The most prominent magnetohydrodynamical (MHD) jet launching mechanisms \citep{blanzn77,blanpa82} rely on having a large poloidal magnetic field in the inner accretion flow. It remains uncertain how jets are produced, but both analytical \citep[e.g.][]{liviet99,meie01} and numerical \citep[e.g.][]{becket08} works favor a strong jet only in the case where the accretion flow has a large scale height magnetic field. In the soft state, a geometrically thin disk exists \citep{shaksu73}, which may \citep{meie01} or may not \citep[e.g.][]{banepu06} suppress any large scale vertical field. However, in some analytical works \citep[e.g.][]{ferr97}, supported by the results of simulations \citep[e.g.][]{casske02}, a thin disk is \emph{preferred} for the production of powerful jets because the radial magnetic tension overcomes the rotation at the disk surface when the disk becomes too thick. As well as soft states, these thin disks may be present in hard states too, and some simulations can reproduce the observed accretion-ejection behaviour of BHXBs in hard states \citep[e.g.][]{petret10}. It has been suggested for example, that the vertical field may be quenched by a changing ratio between the magnetic pressure and the gas and radiation pressure \citep{petret08}. However, the question remains unanswered whether a faint jet is produced in the soft state of BHXBs.

This picture can be tested with a deep radio observation of a BHXB in the soft state. So far such studies have proved inconclusive. Radio emission is usually not detected during the soft state, but any detections could originate in discrete jet ejections launched at previous epochs before the soft state (`relic' jets, which are optically thin and in many cases are directly resolved), or from jet--ISM interactions downstream in the flow \citep{fendet09}. One potential exception could be recent evidence for a compact jet on VLBI-scales in Cygnus X--1 \citep[which in this case is inconsistent with optically thin residual ejecta;][]{rushet11} in a state which could either be soft or soft--intermediate. The question of whether jets are launched in the soft state remains unanswered to date, but if answered, the community can finally define what local physical conditions close to the BH are required to launch relativistic jets. Is a radiatively inefficient accretion flow or a geometrically thick accretion flow, with a scale height similar to the inner radius, required?

\source\ is an X-ray binary discovered in the 1970s \citep*{giacet74} and has been persistently active for at least the last 30 years. The source almost certainly harbors a black hole primary as shown from its X-ray temporal and spectral characteristics, and may contain a rapidly spinning black hole \citep[][see discussion in Section \ref{discussion}]{wijnet02,nowaet08}. There are very few known persistent low-mass X-ray binaries hosting a BH (e.g. \citealt{nowaet08}). In addition, unlike any other Galactic BHXB, \source\ also peculiarly remains persistently in the soft state; although it does fluctuate in hardness and intensity it never reaches the canonical hard state of BHs \citep[e.g.][]{wijnet02}. Only one other BHXB is always observed in a soft state, LMC X--1, a HMXB which is very distant ($> 50$ kpc, in the Large Magellanic Cloud) and therefore much fainter than \source. Since \source\ has not been seen to make a transition to or from the hard state, it will not have launched discrete jet ejections recently, so any radio emission detected from \source\ will be from the core, compact, soft state radio jet. The only reported radio upper limit for \source\ to date is an unconstraining  $< 2.1$ mJy at 5 GHz \citep{nelssp88}.

Here we present new, ultradeep radio observations of \source\ with the Expanded Very Large Array, taken to test this jet quenching paradigm. We make use of the new wide bandwidth capabilities of the instrument to reach $> 2$ orders of magnitude deeper than the deepest existing observation. Multiwavelength (X-ray, UV and optical) data are used to confirm the source is in the soft state during this epoch and we present the broadband spectral energy distribution (SED). We compare our results to detections of jets in the soft state of neutron stars (NSs) and AGN.

\section{Observations}

\subsection{EVLA}
\label{sec:evla}

We made deep observations of \source\ using the wide bandwidth capability of the new Expanded Very Large Array \citep[EVLA;][]{perlet11}.  One hour of test observations were taken on 2010-11-04, followed by a deeper 3-h observation on 2010-12-09 with the same observational setup.  A log of all observations is presented in Table 1. These gave on-source times of 26.9\,min and 121.2\,min, respectively.  We used the wideband 4--8\,GHz receiver system, centering the two 1024-MHz basebands at frequencies of 5.38 and 6.80\,GHz, to provide the most contiguous frequency coverage possible while avoiding the radio frequency interference (RFI) known to exist between 5.93 and 6.27\,GHz.  Each 1024-MHz baseband was comprised of eight 128-MHz sub-bands, each of which contained 64 spectral channels of width 2\,MHz.  The array was in the relatively compact `C' configuration, with a maximum baseline of 3.4\,km.  The data were taken with an integration time of 5\,s.

Data reduction was carried out using the Common Astronomy Software Application \citep[CASA;][]{McM07}.  Bad data arising from shadowing, instrumental issues or RFI were edited out before beginning the calibration.  Notable RFI signals were present at 6.616 and 6.774\,GHz, and the data were Hanning smoothed before further processing to minimize the effects of this interference on surrounding frequency channels.  Bandpass and flux density calibration were carried out using 3C\,286, setting the flux scale according to the coefficients derived at the EVLA by NRAO staff in 2010.  Amplitude and phase gains were derived for both 3C\,286 and the phase calibrator source J1950+0807, located 4.3$^{\circ}$ from \source.  Finally, the calibration was applied to the target source.  Following frequency-averaging by a factor of 8, the two data sets were concatenated, providing a total of 136\,min of on-source integration time after editing out the start of each scan.  The data were then subjected to several rounds of imaging and phase-only self-calibration, using natural weighting during the imaging process for maximum sensitivity.  To allow for the frequency-dependent primary beam size and the varying spectral indices of the sources in the field, deconvolution was carried out using the implementation of the multi-frequency synthesis algorithm of \citet{Sau94} within CASA.

\begin{figure}
\centering
\includegraphics[width=\columnwidth]{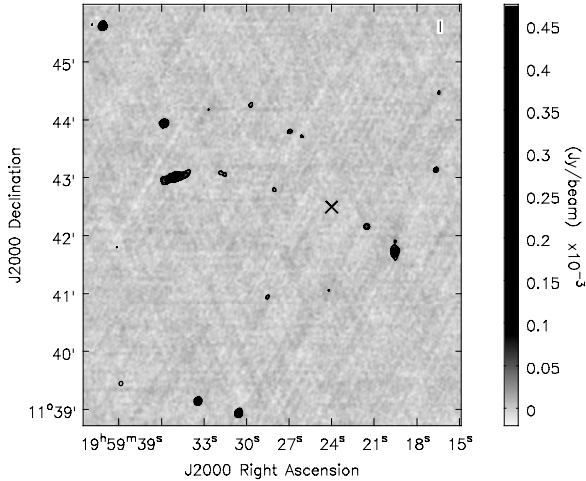}
\caption{Stacked EVLA 6.09\,GHz image of the field containing \source.  Contour levels are at $\pm(2)^n$ times the $5\sigma$ level of 0.0189\,mJy\perbeam, where $n=0,1,2,...$  The best optical position of \source\ is marked with a cross.  While there are a number of background sources in the field, there is no radio emission at the position of the X-ray binary.}
\label{fig:evla}
\end{figure}

No radio source was detected at the best-known optical position of \source\ derived from DSS images (using a combination of USNO-B1, GSC 2.3 and 2MASS), 19$^{\rm h}$59$^{\rm m}$24\fs0, $+11$\degr42\arcmin29\farcs8 (J\,2000).  The $3\sigma$ upper limit on the radio flux density was 11.4\,$\mu$Jy\,beam$^{-1}$.  The best radio image of the field containing \source\ is shown in Fig.~\ref{fig:evla}.

\subsection{Swift} \label{swiftdr}

Swift pointed observations were made on four dates in 2010 November, in between the two radio observations (Table 1).
Swift XRT observations were carried out in Windowed Timing 
(WT) mode. We extracted light curves and spectra using
the Swift-XRT data products generator, following the procedures in
\cite{evanet09}. With each observation having approximately 30000
counts, we grouped each of the source spectra with 150 counts per
spectral bin. The spectral analysis was performed using the standard
HEAsoft X-ray spectral fitting package \small XSPEC \normalsize version 12.6.0. All spectra were fitted in the 0.3--10 keV band, with 1--10 keV fluxes
computed from the fits. We found that all spectra are well described
by a disk blackbody model with absorption. The fitted column densities
are slightly lower than or about the Galactic value, and the inner disk
temperatures vary between 1.3 and 1.5 keV. The mean flux from the four observations is $8.8 \times 10^{-10}$ erg s$^{-1}$ cm$^{-2}$ (1--10 keV), this varies by $< 14$\% between dates.
While the fits are not
formally good, and hence strong claims should not be inferred from
spectral fitting, the strong residuals preventing convergence to
$\chi^2/\nu=1$ are at energies with known response matrix features
(oxygen, gold, and silicon edges), and adding systematic errors of
5--10\% to the data results in good fits. We can reliably infer that there are no strong nonthermal components in the spectrum.

Swift UltraViolet/Optical Telescope (UVOT) data were taken in all six available optical/UV filters. For each filter on each date, the individual images were summed using {\tt uvotimsum} and photometry of the source was performed with {\tt uvotsource}, using an extraction region of radius 3.5\arcsec\ for $v$, $b$ and $u$ filters and 6.0\arcsec\ for $uvw1$, $uvm2$ and $uvw2$ due to a larger point spread function in the latter three UV images. The differences between using 3.5\arcsec\ and 6.0\arcsec\ are 0.01 mag in $uvw1$, 0.11 mag in $uvm2$ and 0.05 mag in $uvw2$. \source\ was detected (at the $\geq 4 \sigma$ level) in all six filters on all four dates. Apparent magnitudes (not de-reddened) are given in Table 2. \cite{marget78} estimated A$_{\rm V}=0.93$ for the interstellar extinction, but the neutral hydrogen column through the whole Galaxy in the direction of \source\ is $1.17 \times 10^{21}$ cm$^{-2}$ \citep{kalbet05} which corresponds \citep{predet95} to A$_{\rm V} \approx 0.65$. We take the range from these two estimates, A$_{\rm V} = 0.79 \pm 0.14$ and use the extinction curve of \cite*{cardet89} applied to the central wavelengths of each UVOT filter \citep{kataet08} to derive the intrinsic, de-reddened flux densities.

\subsection{Faulkes Telescope North}

Imaging of \source\ was performed in four optical filters (Bessell $B$, $V$, $R$ and Sloan Digital Sky Survey $i'$) with the robotic, 2-m Faulkes Telescope North (FTN) at Haleakala on Maui, USA on 2010-11-23. The airmass of the target was 1.3--1.4 and the conditions were good, with a seeing of 1.1''. The EM01 camera was used, with a field of view of 4.7 $\times$ 4.7 arcmin. Reduction, photometry and flux calibration were performed as described in \citeauthor{russet10} (2010; see above for details of the de-reddening procedure). To calibrate the Faulkes B-band data we measured the UVOT $b$-band magnitudes of the two field stars used in \cite{russet10} using the same method as for \source\ (see Section \ref{swiftdr}). We find $B \sim b = 17.13 \pm 0.08$ for star 1 (at $19^h 59^m 22^s.5$ $+11^{\circ}42' 20''$ J2000) and $B \sim b = 17.13 \pm 0.08$ for star 2 \citep[at $19^h 59^m 23^s.4$ $+11^{\circ}42' 06''$ J2000; the differences between Bessell $B$ and UVOT $b$ magnitudes are less than these errors; see][]{poolet08}. The UVOT $v$-band magnitudes of the two field stars also agree with the values derived in \cite{russet10} to an accuracy of $0.05 \pm 0.08$ mag. For \source\ we find $B = 18.84 \pm 0.12$; $V = 18.74 \pm 0.11$; $R = 18.60 \pm 0.11$; $i' = 18.71 \pm 0.05$ on 2010-11-23.





\begin{table}
\begin{center}
\caption{UVOT optical/UV magnitudes of \source.}
\begin{tabular}{lllll}
\tableline\tableline
Filter & \multicolumn{4}{c}{Date in 2010 November} \\
       & 16 & 19 & 22 & 25 \\
\tableline
$v$    & 18.79(26) & 18.38(20) & 18.71(26) & 18.90(31) \\
$b$    & 18.81(15) & 18.60(13) & 19.01(17) & 19.26(21) \\
$u$    & 17.50(10) & 17.67(11) & 18.00(13) & 17.81(12) \\
$uvw1$ & 17.49(11) & 17.52(12) & 17.76(13) & 17.88(14) \\
$uvm2$ & 17.66(12) & 17.77(13) & 17.82(13) & 18.31(17) \\
$uvw2$ & 17.75(9)  & 17.74(10) & 17.83(10) & 18.10(11) \\
\tableline
\end{tabular}
\end{center}
\end{table}

\section{Results and analysis}

\subsection{The soft state jet quenching factor}

The EVLA observations have provided an ultradeep upper limit to the radio flux of \source. The Swift XRT spectral and variability properties confirm the source was still in the soft state during this epoch, as expected (see Section 2.2). We can therefore estimate the soft state radio jet quenching factor by plotting our data on the well known radio--X-ray luminosity diagram of hard state BHXBs \citep[Fig. 2; data are from][]{fendet10,calvet10}. Given the uncertainty in its distance \citep[$7 \leq$ d/kpc $\leq 22$;][]{marget78,nowaet08}, \source\ is at least 330--810 times fainter than the BHXBs that follow the established hard state radio--X-ray correlation \citep{corbet00,gallet03} at the same (1--10 keV) X-ray luminosity. In addition, the faintest of the `radio-quiet' population of hard state BHXBs \citep[see e.g.][]{calvet10} is H1743--322, which is $\geq 14$ times more luminous than \source.

\subsection{The broadband SED}

In Fig. 3 we present the first broadband, radio--X-ray SED of \source. The whole SED can be explained by thermal emission from the accretion disk. The X-ray spectrum is soft, with 1.3--1.5 keV for the inner disk blackbody temperature, and the optical/UV SED is blue, with a spectral index of $\alpha = +0.50 \pm 0.31$ (where $F_{\nu} \propto \nu^{\alpha}$) between 1928\AA\ ($uvw2$) and 7545\AA\ (i') taking into account the uncertainty in the extinction. This is typical for a bright BHXB in which the (possibly irradiated) outer disk dominates these wavebands. The radio-to-optical spectral index is $\alpha > +0.24$, and no red excess is seen in the optical/UV SED, confirming a negligible synchrotron jet contribution also at these higher frequencies.
The optical V,R,i' magnitudes during this epoch are slightly brighter than (but well within the range of) the mean magnitude measured from long-term optical monitoring in 2006--2009 \citep{russet10}. Both the X-ray and optical properties during this epoch are typical for the source, confirming that \source\ remained in the soft state.

\begin{figure}
\includegraphics[width=6.1cm,angle=270]{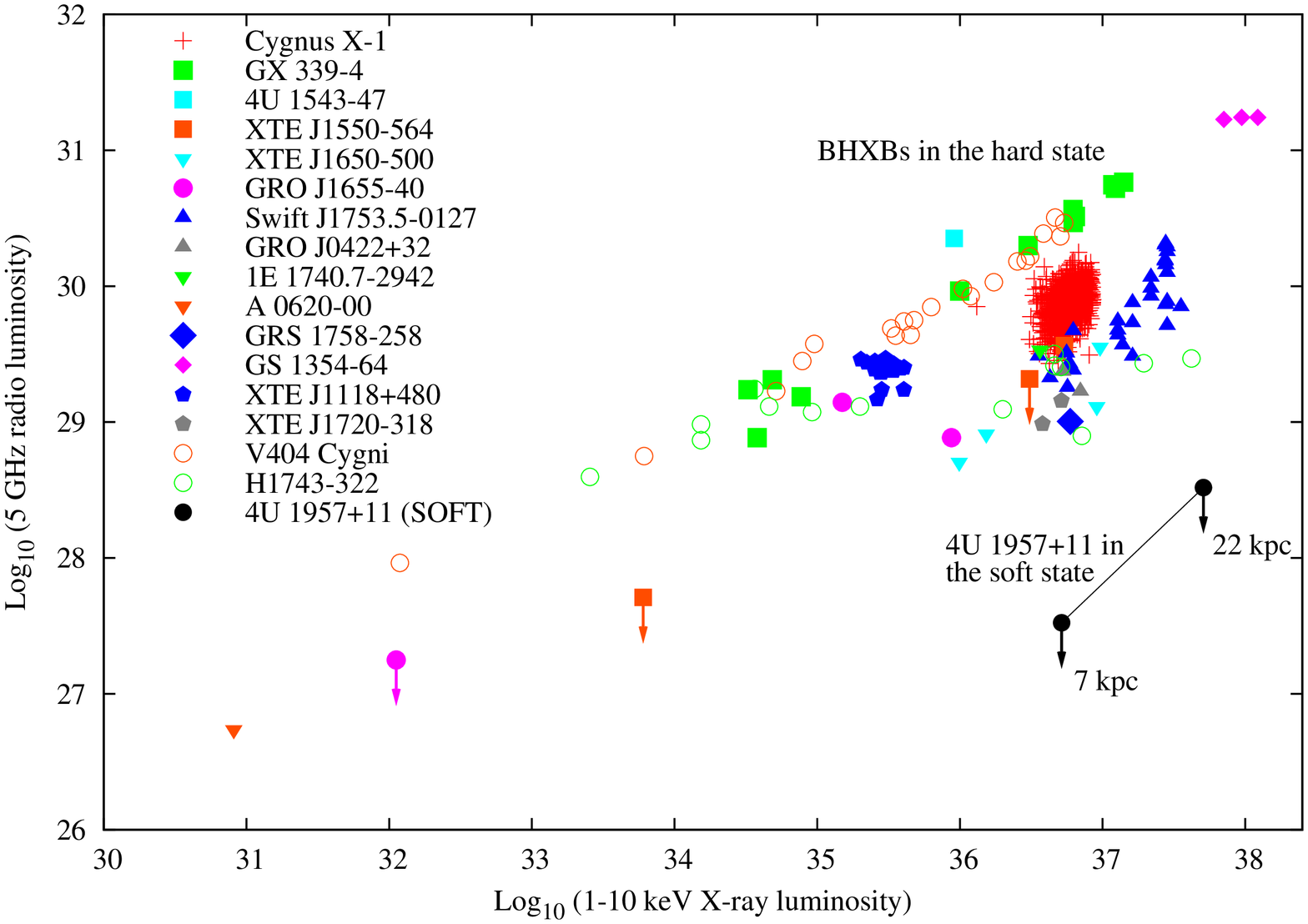}
\caption{Radio -- X-ray luminosity diagram of hard state BHXBs \citep[an updated compilation, from][]{fendet10,calvet10} with our \source\ soft state data. 
}
\end{figure}

\section{Discussion} \label{discussion}

We have shown that the radio flux of \source\ in the soft state is suppressed by a factor $> 330$ compared to the hard state radio--X-ray correlation of BHXBs. Dynamical mass estimates of \source\ have not been attempted because it has never faded to quiescence, so is \source\ definitely a BHXB? As well as an X-ray spectrum typical for a soft state BHXB, \source\ displays X-ray timing properties similar to other BHXBs in the soft state, and the relation between hardness and luminosity would be unusual for a neutron star XB \citep{wijnet02,nowaet08}. Its overall X-ray behaviour is similar to the persistent source LMC X--3, which contains a $> 5.8$ M$_{\odot}$ black hole \citep[see ][and references therein]{wijnet02}. The X-ray spectral properties could only be like those of a NS if the source is accreting at close to the Eddington luminosity (but if this were the case the level of variability is too low for this class of NS), i.e. at a very large distance \citep{wijnet02}. It was also shown from the optical--X-ray ratio that the system is only likely to harbor a neutron star if the distance is small \citep{russet10}, so at any distance a neutron star accretor is not favored. Thus \source\ is far more likely to be a black hole than a neutron star.

Soft state radio detections from transient BHXBs are typically much brighter than our ultradeep upper limits for \source\ \citep{fendet09}, but very likely originate in discrete ejecta launched previously over the transition to the soft state \citep[at the `jet line'; see][]{fendet09}. Being persistently in the soft state, the radio emission of \source\ is not confused by discrete ejecta. \cite{coriet11} report some deep radio upper limits for the BHXB H1743--322 during a soft state, one of which, $< 30 \mu$Jy, is $\sim 700$ times fainter than its radio luminosity at the same X-ray luminosity in the hard state. Clearly, a jet model is required that satisfies the condition of a soft state jet quenching factor of almost three orders of magnitude.

It has been argued that the geometrically thin disks in soft state BHXBs may have much weaker vertical fields than hard state disks, and a jet quenching factor of $\sim 100$--200 was estimated by \citeauthor{meie01} (2001; this value has a dependence on the disk viscosity parameter and the BH spin).  Alternatively, the jet power may not depend strongly on the disk thickness. A large scale height magnetic field may persist and thread the disk in the soft state, but a change in disk magnetization could prevent a steady-state jet from being launched \citep{petret08}.

\begin{figure}
\includegraphics[width=6cm,angle=270]{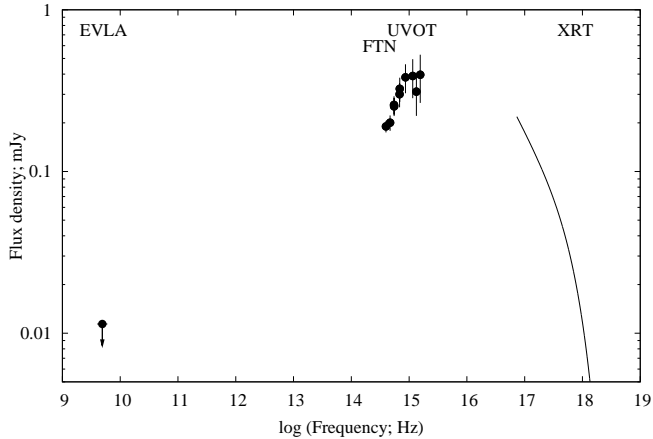}
\caption{Unabsorbed, broadband radio to X-ray SED of \source. EVLA (5--7 GHz), FTN (i', R, V, B-bands) optical, Swift UVOT (v, b, u, uw1, um2, uw2) optical/UV and Swift XRT (0.3--10 keV model fit from the first epoch) X-ray data are plotted.}
\end{figure}

While our results imply dramatic jet suppression in the soft state of BHXBs, this is not usually the case for accreting NSs. The stringent upper limit therefore also favors a BH primary in \source. Radio emission in soft state NSs, when detections are made, are generally fainter than expected from their radio--X-ray luminosity correlations as measured in hard states, and are sometimes consistent with being compact, steady jets like those of hard state BHXBs \citep[e.g.][and references therein]{millet10}. The boundary layer on the NS surface may provide a differentially rotating region from which jets can be launched for soft state NSs, but not soft state BHs \citep{livi99,macc08}.

Radio emission has been detected from AGN in a state analogous to the soft state of BHXBs. These are AGN with evidence for thin disks (mostly Seyferts), and are identified by their power spectra or sometimes their spectral energy distributions. Some of the least luminous Seyfert radio cores \citep{giropa09,joneet11} have radio luminosities $\sim 10$--100 times fainter than the fundamental plane relation of \cite{gultet09}. This suggests jet production does not turn off entirely for soft state AGN, and in fact 
jet quenching factors in different AGN may cover a range $\sim 10$--1000 \citep[e.g.][]{maccet03,sikoet07}. Deeper radio observations of soft state BHXBs are required to test whether stellar-mass BHs also have such a large range of jet quenching factors. Ideal candidates would be nearby BHXBs such as A0620--00 or XTE J1118+480; deep, high resolution radio observations of the core during a soft state would probe jet quenching factors of $> 10^3$. The two with deep upper limits so far, \source\ and H1743--322 \citep{coriet11}, imply soft state BHXBs may be more radio faint, on average, than soft state AGN, which hints toward a fundamental difference in the jet launching process in BHs of different masses surrounded by geometrically thin disks.

\acknowledgments

The Faulkes Telescope North is maintained and operated by Las Cumbres Observatory Global Telescope Network. We thank the European Union for funding under Initial Training Network 215212: Black Hole Universe. DMR acknowledges support from a Netherlands Organisation for Scientific Research (NWO) Veni Fellowship. FL acknowledges support from the Dill Faulkes Educational Trust.



{\it Facilities:} \facility{EVLA}, \facility{Faulkes Telescope}, \facility{Swift}.

\clearpage

\end{document}